\documentclass[12pt]{article}
\usepackage{amsmath,amssymb,stmaryrd,cite}
\usepackage{graphicx}

\newcommand{\ba}{\begin{array}}
\newcommand{\ea}{\end{array}}
\relax

\begin{document}
\thispagestyle{empty}

\title{Range of stability of solar neutrino flux from the SAGE experiment data}

\author{{\bf V.~A.~Koutvitsky,} {\bf V.~B.~Semikoz}\thanks{Correponding auth.: semikoz@yandex.ru},\\
 {\small Pushkov Institute of Terrestrial Magnetism, Ionosphere and Radiowave}\\
 {\small Propagation of the Russian Academy of  Sciences,}\\
 {\small IZMIRAN, Troitsk, Moscow region, 142190, Russia,}\\
\and{\bf D.~D.~Sokoloff}\\
{\small Department of Physics, Moscow State University, 119999, Moscow, Russia} }
\date{\today}

\centerline{\LARGE Range of stability of solar neutrino flux}
\centerline{\LARGE   from the SAGE experiment data}
\vskip 12pt
\centerline{\bf V.~A.~Koutvitsky, V.~B.~Semikoz, and D.~D.~Sokoloff}
\vskip 12pt

  \begin{abstract}
  We study the extent to which the SAGE experiment data indicate the permanence of the solar neutrino flux. It is shown that in the first approximation this flux is constant and its distribution function is unimodal.
  Using a more detailed analysis one finds out that data of the first years of experiment (1990-1992) demonstrate a time dependence which is slightly different from what was found for 
the subsequent years (1993-2006). 
The distinctive feature of the first years of experiment is a high dispersion of neutrino flux in comparison with the following epoch. 
We discuss possible astronomical consequences of this result.

  \end{abstract}

\maketitle

\section{Introduction}
The problem of a possible presence of solar neutrino flux
periodicities in different neutrino experiments and
the possibility to find correlations of flux changes with dynamics
of different tracers of solar activity attracts permanent 
attention of researchers. For instance, in the works
\cite{Stozhkov,Obridko} one states the presence of anticorrelation
of the neutrino flux and solar activity while in the paper
\cite{Sturrock_2006} one postulates the existence of rotational
modulation of neutrino flux with the period $\sim$ 30-60 days. 
In the first case, in the paper \cite{Obridko}, the Homestake experiment data 
during 1970-1990 years \cite{Homestake} are confronted with 					
variations of the solar surface magnetic fields. In the second case,
GALLEX experiment data for 1992-1997 are analyzed. In the paper
\cite{Sturrock_2006} one finds bi-modality of the neutrino flux 
distribution function in the GALLEX experiment. On the other hand,
in the paper \cite{Pandola}, the author could not find variations of the
neutrino flux in the data of GALLEX-GNO experiments.

It is difficult to follow a relation of  {\it the Sun short-periodic
rotation} with the variations of solar neutrino flux \cite{Sturrock_2006} 
using mechanism of the neutrino resonant spin-flavor precession (RSFP) 
such as (i) conversion of the left-handed electron neutrino into 
the sterile right-handed one,$\nu_{eL}\to \nu_{sR}$, due to 
a large neutrino transition magnetic moment ($\mu_{\nu}\sim 10^{-11}\mu_B$) and 
(ii) {\it slowly changing} magnetic field in the convective zone (CZ) of the Sun
\cite{Pulido}. The variations of magnetic field are related with
rather longer periods of the order of solar cycle (see details below
in Section IV).

In connection to the above it seems to be useful our returning to
the question how much (in what measure) the SAGE experiment data
allow to select variations of the solar neutrino flux. Moreover, if
such variations are noticeable with a some probability in 
observations then which physical mechanisms can (or can not) govern
such flux variations.

Carrying out the analysis of the SAGE experiment data we rely here
on the solar neutrino data in that experiment for 1990-1997
\cite{Gavrin2}, 1990-2001 \cite{Gavrin3} and also the data of the
recent measurements until 2006 year \cite{Gavrin1} that allow us to
analyze all neutrino events for the period 1990-2006
\footnote{V.N. Gavrin, private communication. Authors thank V.N.
Gavrin and Bruce Cleveland who put their materials at our disposal
(including data for 2006 year) and acknowledge useful discussions
with them. In the paper \cite{Gavrin1} the results of the global
analysis of the registered solar neutrino fluxes are given using the
method of the maximum likelihood, including SAGE data until December
2005.} (in Fig. 1), i.e. during the longest period of solar neutrino
observations after the chlorine-argon (Homestake) experiment
\cite{Cleveland} accounting for the calibration of detector in the
SAGE experiment \cite{Gavrin4}.
\begin{figure}[htb]\label{Fig1}
{
\begin{center}
\includegraphics[width=0.85\textwidth]{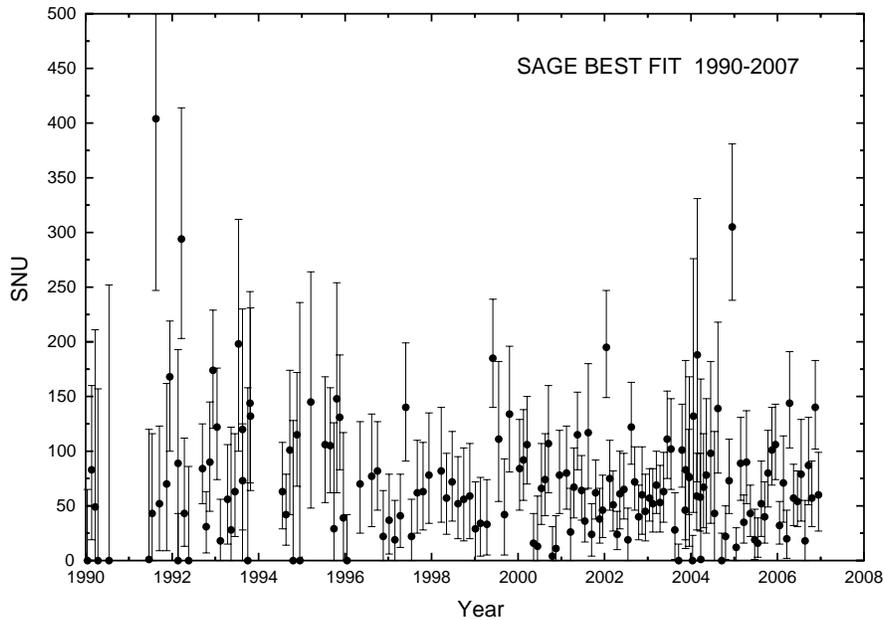}
\end{center}
} \caption{Time series for SAGE solar neutrino data for 1990-2007.
Points are best fits of SNU. Error bars for best fits are shown.}
\end{figure}

Carrying out our statistical analysis we take into account the fact
that duration of one run in the SAGE experiment is $\sim 30$
days that excludes selection of short-periodic processes. On the
other hand, we have no reliable theoretical predictions of which kind
of the neutrino flux changeability (periodicity, pulsed processes or
something else) one could expect for given experiment data.
Therefore, in the statistical analysis of the SAGE experiment data we
try to use general statistical tests not requiring any model of expected
variations specified by a certain dependence on unknown parameters
(e.g. by frequencies of variations).

We do not discuss here neutrino flux variations happening outside
the Sun. These are day/night variations arising due to regeneration
of electron neutrinos traveling (in night) through the Earth, and
seasonal variations appearing due to geometric effects such as $\sim
L^{-2}$ from the Earth orbit eccentricity ($\varepsilon=0.0167$)
always taken into account in all radiochemical experiments in the
correspondence with a data (month) of measurements. Finally nadir
angle and latitude of a detector for events registered are taken
into account (see e.g. in \cite{Lisi}). Here we discuss the
possibility for more slower variations of neutrino flux using the
SAGE experiment data and study their relation with phenomena inside
the Sun that have periods $\sim$ years and much more.

The paper is organized as follows: in Section 2 we present SAGE
solar neutrino datasets and give the simplest
statistical analysis of the SAGE datasets. From the statistical
analysis of the SAGE solar neutrino data we find in Section 3 two epochs,
1990-1992, and 1993-2006, which have some slightly distinguishable
capture rate profiles which could be associated with some kind of
the varying solar activity. In Section 4 we discuss possible
mechanisms responsible for the variation of the solar neutrino data
and stress their imperfection from the point of view of particle
physics and solar physics. In the last Section 5 we discuss results of
our analysis confronting them with the mechanisms discussed in the 
previous Section.

\section{Statistical analysis of neutrino flux}
Below we perform a statistical analysis of the SAGE data in order to
address the following problems: (i) to what extent the data can be
considered as a realization of a stationary random process, e.g.
Poisson process; (ii) if this is the case, what is the probability 
distribution function
(pdf) of that random process? Let us note that if a random process
is essentially unsteady then data of one time-series do not allow to
restore a distribution function because it changes essentially over
time and we dispose at each time moment only one measurement to
determine pdf.

We address the first question using a simple test presented in
Fig.~\ref{Fig3}. If the time series presented in Fig.~\ref{Fig1} can
be considered as a realization of a stationary random process $f$
with nonvanishing mean value $<f>$ then the cumulative capture rate
$g(n)=\Sigma_1^n f(n)$  summing over the exposition run number $n$
has to grow with $n$ as $<f>n + \dots$ where the symbol $\dots$ mean
terms which grow slower than $n$. If the plot does not demonstrate a
linear shape we have to reject the hypothesis that $f$ is a
stationary random process. Note that if there are some gaps in
time-series we have just to omit corresponding months in the
calculation of the exposure run numbers. The plot of $g(n)$ as it is
obtained in accordance with the SAGE data is presented by solid line
in Fig.~\ref{Fig3}. We see that the plot is amazingly linear for $n
> N \approx 30$. Its slope gives $<f>\approx {\rm const}$ for the epoch
after 1993.

Hence not considering small deflections from the linear law during
first 2-3 years of experiment we have no grounds to speak about any
variations of neutrino flux (i.e. about deflections from the linear
law given by the function $g(n)$), in any case  at the time scales
acceptable for analysis of the given experiment.

\begin{figure}[htb]\label{Fig2}
{
\begin{center}
\includegraphics[width=0.8\textwidth]{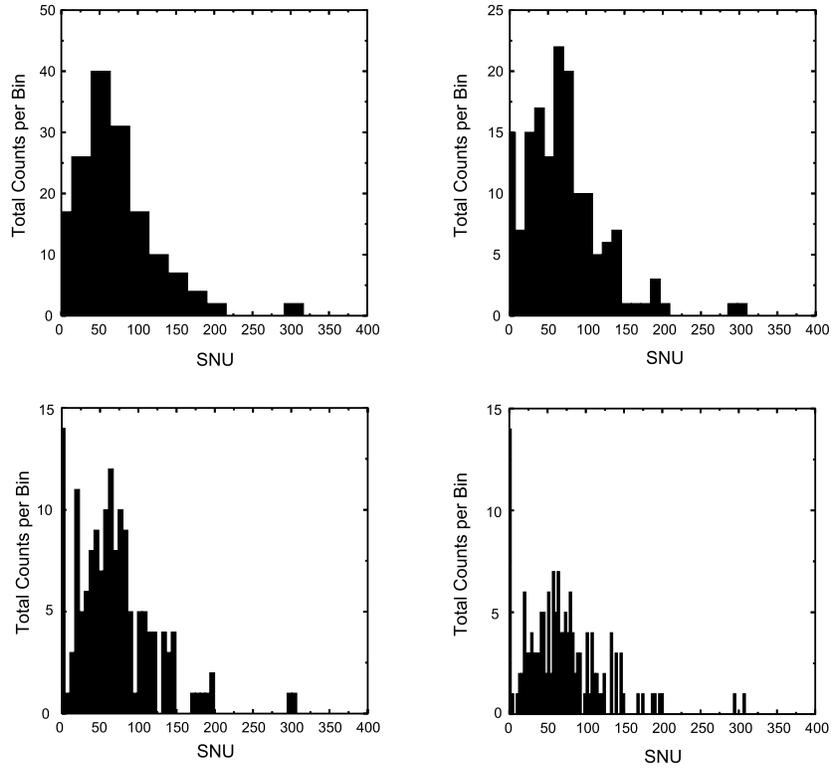}
\end{center}
}
\caption{The histograms of the SAGE data %
grouped by SNU value into 16, 32, 64, and 128 bins.}
\end{figure}

Then we can plot the distribution function of the studied random
process which is considered as a steady one. Since the volume of
analyzed sampling is small we do not rely on evaluation of the probability
density based on histograms that show the relative number of
exposures corresponding to a given interval of neutrino fluxes.
Instead of that we plot the corresponding integral value,  namely
empirical distribution function, i.e.  the relative exposure number
with the capture number less than the given one (in Fig.~{4}). 
As any distribution function it vanishes for small
$x$ (in fact, at $x=0$) and tends to the unity for large $x$. In the
given case the obtained dependence changes smoothly between  those
limiting values. This pdf is similar to the gaussian distribution
shown in the same Figure (panel b). Let us note that such pdf can not be
exactly gaussian since it is not negative due to its physical sense.
As a whole Fig.~{4}) demonstrates unimodal distribution of
the neutrino flux. The same conclusion follows from the analysis of
histograms for distribution density comprising all experiment 
period (see in Fig.~\ref{Fig2}). One can easily see that after a reasonable 
bunching of data (over 16 bins) the corresponding histogram  
has no features of bimodality.
The more detailed histograms shown in Fig.~(\ref{Fig2}, b,c,d)
demonstrate only significant dispersion of experimental data and
also can not be as indication a bimodality of the considered
distribution.

\section{Statistical properties of SAGE experiment data during 1990-1993}
\begin{figure}[htb]\label{Fig3}
{
\begin{center}
\includegraphics[width=0.7\textwidth]{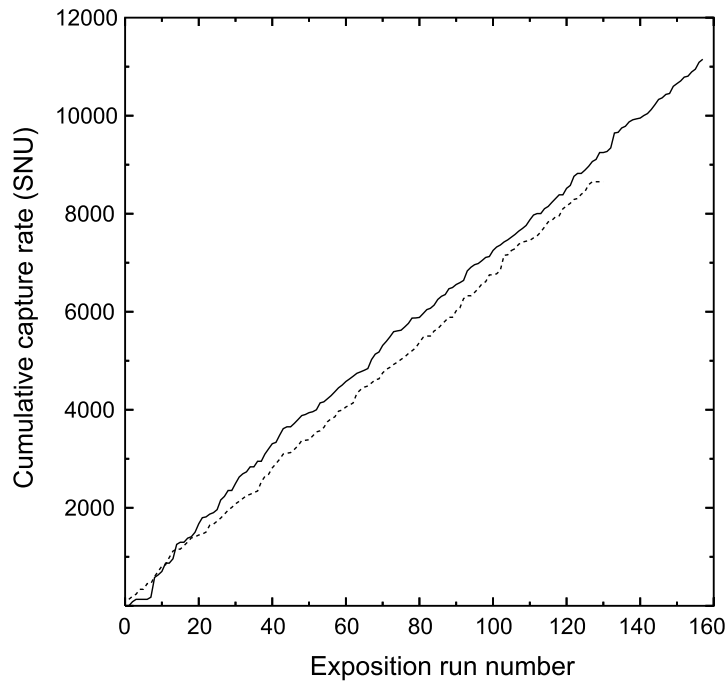}
\end{center}
} \caption{Cumulative capture rate (SNU) versus the exposition run
number (solid line). Dashed line presents the cumulative capture
rate versus the exposition run number for the last part of the SAGE
experiment (after the year 1993; dashed line).}
\end{figure}

Let us note that data of the first 30 exposure runs precipitate from
the description given above as it is seen in the beginning of curve
shown by the solid line in Fig~\ref{Fig3}. In order to demonstrate
that the 30 exposure runs are sufficient to isolate a linear growth
we perform the same test omitting first 30 exposure runs (dashed
line in Fig.~\ref{Fig3}). We see that the line obtained demonstrates
a linear growth almost from its beginning and the slope is the same
as for the solid line.

We conclude also that the probability distribution function (pdf) of the random
process $f$ has to be considered for epoches I and II separately.
We present corresponding data in Fig.~4 for epoch I by
dashed line and for epoch II by solid line. For that summing over exposure run
numbers with a given rate (SNU) and normalizing partial sums on the run numbers (=30)
during epoch I and (=127) during epoch II we get corresponding pdf curves.

We see that pdf for epoch I was much wider than for epoch II. 
The mean value $<f>$ for epoch I is slightly larger than the
mean value for epoch II. This means that during epoch I a noticeably
larger dispersion of the neutrino flux was registered than during
epoch II. The distribution functions for both epochs indicate a
unimodal distribution while these distributions are different. Let
us stress that data under analysis are not sufficient to insist on a
stability of random process considered during the epoch I, i.e., the
interpretation of the corresponding empirical  distribution function
remains to some extent as only conventional.

\begin{figure}[htb]\label{Fig4}
{
\begin{center}
\includegraphics[width=1.0\textwidth]{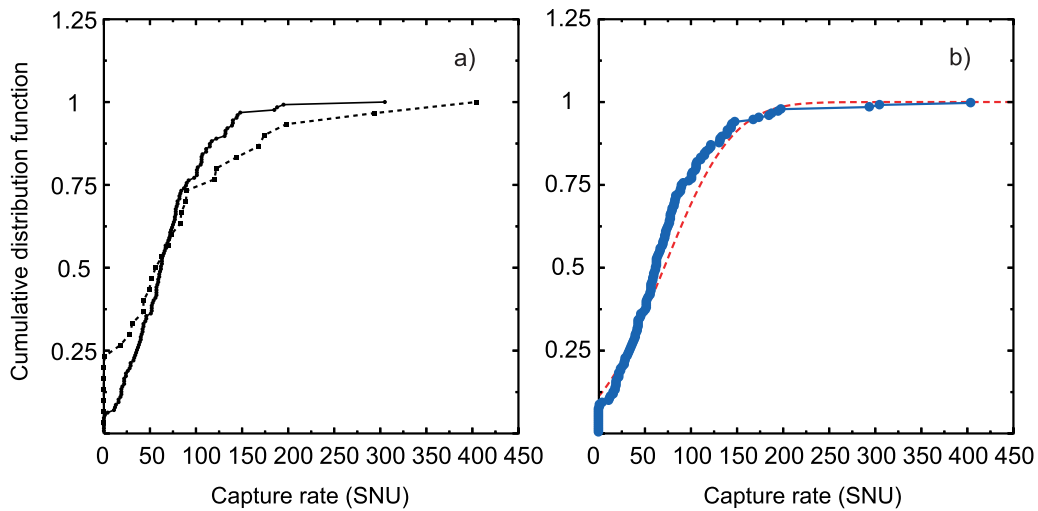}
\end{center}
} \caption{Cumulative distribution functions (probability, i.e.
relative number, to obtain capture rate less than $x$ SNU) versus $x$: 
a) - for epoch I (dashed line) and epoch II (solid line); 
b) - for all SAGE data (solid line), gaussian distribution (dashed line).}
\end{figure}

The assumption that the difference between epochs I and II corresponds to some real
processes in the Sun  leads immediately to the serious astronomical
conclusions. In the next Section we show that it is difficult to
interpret this distinction through processes occurring within the
solar convective zone.

\section{Solar magnetic fields and possible mechanisms responsible for solar neutrino flux variation }

The solar activity is connected with changing magnetic fields within
solar interior, e.g. in terms of the dynamo theory of the solar
cycle \cite{Sokoloff}. Some traces of these changing magnetic fields
are observed in the bipolar active regions consisting of sunspots as
bundles of magnetic loops floating upwards from an initially
horizontal magnetic field in the convective zone (CZ). The polarity
rules of the active regions on the sun show: (i) that the horizontal
magnetic field below the surface is nearly East-West oriented, (ii)
that the toroidal field ${\bf B}_{tor}$ direction is opposite in
each hemisphere, and (iii) that a polarity reversal takes place from
one cycle to the next. Though magnetic fields in sunspots (surface
fields $B_s$) were discovered long time ago we do not know how they
change in time and how these variations are related to the
large-scale (toroidal $B_{tor}$) magnetic field at the bottom of CZ.
Moreover, $B_s$-measurements do not provide magnetic field at each
point in the Sun (and, which is still more important, not the field
on the "solar center-observer" line), but rather an uncompensated
part of the general field whose relation to $B_{\perp}$ influencing
neutrino spin-flip is unknown. Thus, these surface magnetic field
data (as well as the Wolf numbers) can be used only as indirect
information on $B_{\perp}$ and CZ magnetic fields.

\subsection{Spin-flavor precession (SFP) scenario of the neutrino flux variation}

In both cases of long time neutrino datasets mentioned above the authors 
\cite{Obridko,Sturrock_2006} assumed the presence of a non-zero neutrino  
magnetic moment $\mu_{\nu}\neq 0$ due to which some part of the electron  
neutrino flux can be converted in the changing solar magnetic fields to 
another neutrino species not registered in radiochemical 
experiments which are sensitive to the charge current interactions 
provided by the left-handed electron neutrinos $\nu_{eL}$.

For instance, this process can be an efficient (vacuum non-resonant)
active-active Majorana neutrino spin-flavor precession
$\nu_{eL}\to\bar{\nu}_{aR}$, $\nu_{a L}\to \bar{\nu}_{eR}$ in random
magnetic fields within the diluted solar CZ that happens after the
dominant LMA MSW conversion $\nu_{eL}\to \nu_{aL}$ in dense matter
of the radiative zone (RZ)\cite{Rez} for which the LMA neutrino
mixing parameters at $1~\sigma$ ($3\sigma$) level \cite{Concha},

\begin{equation}\label{LMA}\Delta
m^2_{21}=7.67^{+0.22}_{-0.21}\left(^{+0.67}_{-0.61}\right)\times
10^{-5}~eV^2,~~~~~~~~\theta_{12}=34.5\pm
1.4\left(^{+4.8}_{-4.0}\right),\end{equation}
are firmly established
from all neutrino experiments including KamLAND with reactor
antineutrinos $\bar{\nu}_{e}$ \cite{KamLAND}. There  remains an
open problem for the scenario with the {\it rms} magnetic field
$b(t)=\sqrt{\bar{b^2}}$, namely, how it depends on time during solar
activity to be relevant for our discussion of varying neutrino
fluxes.

Another possibility is a more speculative RSFP conversion to an
additional sterile neutrino in a regular CZ magnetic field with the
appropriate $\Delta m^2_{10}=O(10^{-8})~eV^2$ that proceeds after
the dominant LMA MSW in RZ with the mixing parameters for active
neutrino species given by Eq. (\ref{LMA}) \cite{Pulido}. In addition
to the unknown connection of a strong regular CZ magnetic field
($B_{\perp}\sim 300~kG$) with the measured varying surface magnetic
fields $B_s$ in sunpots ($\sim kG$) the assumption of a light
sterile neutrino together with unknown magnetic moment are too
doubtful.

The present laboratory bounds on the neutrino magnetic moment are
given by the reactor antineutrino scattering off electrons at low
energies as $\mu_{\nu}\leq 9\times 10^{-11}\mu_B$ in the MUNU
experiment \cite{MUNU} and $\mu_{\nu}\leq 5.8\times 10^{-11}\mu_B$
by GEMMA spectrometer \cite{GEMMA} while there are more severe
astrophysical bounds $\mu_{\nu}\leq 3\times 10^{-12}\mu_B$
\cite{Raffelt, Rez}.

\subsection{Parametric resonance of MSW oscillations in the presence
of matter density perturbations in radiative zone (RZ)}

There is another possibility to observe time variations of solar
neutrino fluxes not exploiting idea with neutrino magnetic moment
while relying on the presence of magnetic fields in RZ and commonly
held LMA MSW scenario of neutrino oscillations. This is an old idea
of the parametric resonance for matter density perturbations
influencing MSW oscillations \cite{Krastev} when the wave length
$\lambda_{\delta \rho}$ for the matter perturbation $\delta
\rho(t,r)/\rho$ in the total neutrino potential
$V_{MSW}(1+\delta\rho/\rho)$ entering Schr\"{o}dinger equation that
governs neutrino oscillations coincides with the neutrino
oscillation length, $\lambda_{\delta \rho}\approx l_{\nu}=4\pi
E/\Delta m^2_{12}$. This is impossible for the long p,g-mode waves
in the standard helioseismology \footnote{Note that helioseismology
bound $\delta \rho/\rho <0.01$ for matter density perturbations in
RZ which is coming from the solution of inverse problem using the
data with long wavelength p-modes \cite{Christensen} fails at short
distaces corresponding to the neutrino oscillation length.}, for
which $\lambda_{p,g}\sim 10^4-10^5~km$ is much bigger than
$l_{\nu}\sim 100-200~km$ for $O(MeV)$-neutrinos while this can be
realized for short magneto-gravity (MG) waves in the horizontal RZ
magnetic field, ${\bf B}\perp \nabla \rho$ \cite{Burgess}.

The appearance of the Alfv\'{e}n resonance for such geometry of
magnetic field leads to the rise of many spikes of density
perturbations separated by the distances of the order 100-200 km
exactly as the solar neutrino oscillation length that provides the
parametric resonance \cite{Burgess}. On the other hand, the cavity
for magneto-gravity waves bounded by the center of the Sun and these
spikes blocks g-mode propagation within RZ, or they become
evanescent even deeper than the bottom of CZ.  Since the the
presence of MG resonance tends to decrease MSW effect, the
prediction would be that the observed rate of solar electron
neutrino events is maximized when the Earth is closest to the solar
equatorial plane (December and June) and is minimized when the
furthest from this plane (March and September)\cite{Burgess}. This
happens because of 7-degree inclination of the Earth's orbit
relative the plane of the solar equator in which neutrinos
registered at the Earth permeate the horizontal RZ magnetic field
${\bf B}\perp \nabla \rho$ and these seasonal $\nu$- flux variations
are possible {\it without neutrino magnetic moment}, $\mu_{\nu}=0$.
Other periodicity connected with the low frequency MG waves (periods
of order $\sim$ days, weeks) is less pronounced in \cite{Burgess}
because of poor neutrino event statistics at present.

In addition to, the model of a central (RZ) magnetic field is less
elaborated than models of CZ fields, and, moreover, MG modes if such
form of g-modes exists should be invisible on the solar surface.

Resuming we would like to stress that all known mechanisms for the
variations of electron neutrino flux deep the solar interior rely on
varying magnetic fields which influence neutrino oscillations either
in SFP assuming a nonzero neutrino magnetic moment or due to  the
parametric resonance of LMA MSW oscillations. In both cases there
are still too much uncertainties of a solar MHD model to apply
appropriate issues for neutrino propagation in the Sun.

\section{Conclusions}
Thus, we showed that except of the initial period 1990-1992 the SAGE
experiment data describe the solar neutrino flux as a steady random
process having the simplest one-modal distribution. The peculiarity
of SAGE data for the initial period of observation is not some sort
of periodic process with a period comparable with the whole time of
SAGE experiment. Considering that data separately it is simpler to
explain them as a display of some instability apearing from time to
time in solar CZ. However the conclusion  about the existence of
such instabilities in CZ seems to be too radical and premature being
based on these data only.

This is confirmed by a short analysis of MHD models and mechanisms of neutrino interaction with solar magnetic fields
as given in the previous Section.
 On the first glance, after the maximum of the solar
activity in 1990 CZ magnetic field strengths were decreasing
resulting in an increase of the electron neutrino survival
probability if we rely on the SFP scenario, let us say, for
subdominant $\nu_e\to \nu_s$ conversion in CZ as in the paper
\cite{Pulido}. Thus, SAGE registered more solar electron neutrinos
during epoch I.  However, we do not understand why when 11 years
passed, e.g. in 2003-2006 we do not observe the same increase of the
solar neutrino flux during next 23 cycle of the solar activity with its
minimum somewhere in 2007. This turns us to think that another solar
periodicity (with a longer period $>$ 11 years) acts on solar
neutrino flux during epoch I.

Nevertheless we think that the potential ability of such conclusion itself is enough to proceed SAGE experiment
over a long period of time.
\section{Acknowledgements}
We thank discussions with Vladimir Gavrin and Bruce Cleveland. We thank also 
the Program of RAS Presidium "Solar activity" \# 16 for financial support.
 \eject

\end{document}